\documentclass{aa}
\usepackage{graphicx}
\usepackage{times}
\newcommand{\etal}{{et al.~}}
\newcommand{\Msun}{\>{\rm M_{\odot}}}
\newcommand{\kms}{\>{\rm km}\,{\rm s}^{-1}}
\newcommand{\Mpc}{\>{\rm Mpc}}
\newcommand{\Jybeam}{\>{\rm Jy/\! beam}}

\begin{document}
   \title{Relation between dust and radio luminosity in optically selected early type galaxies}


   \author{D. Krajnovi\'c
          \inst{1}\thanks{on leave of absence from Institute Rudjer Boskovic, Bijanicka cesta 54, 10000 Zagreb, Croatia}
          \and
          W. Jaffe\inst{1}
          }

   \offprints{D. Krajnovi\'c}

   \institute{Sterrewacht Leiden, Postbus 9513, 2300 RA Leiden, 
The Netherlands\\
              \email{davor@strw.leidenuniv.nl}}

\date{Received <> ;  Accepted <>}

\abstract{ 
We have surveyed an optical/IR selected sample of nearby E/S0 galaxies
with and without nuclear dust structures with the VLA at 3.6 cm to a
sensitivity of 100 $\mu$Jy.  We can construct a Radio Luminosity
Function (RLF) of these galaxies to $\sim 10^{19}$ W Hz$^{-1}$ and
find that $\sim 50$\% of these galaxies have AGNs at this level.  The
space density of these AGNs equals that of starburst galaxies at this
luminosity.  Several dust-free galaxies have low luminosity radio
cores, and their RLF is not significantly less than that of the dusty
galaxies.
  
   
 }

\titlerunning{Dust and radio luminosity function in early type galaxies}
\authorrunning{Krajnovi\'c \& Jaffe}

\maketitle
%

\section{Introduction}

 Research conducted during the last decade gave a new look to nearby
elliptical galaxies previously considered as old, uniform systems with
little gas or dust. Images from the Hubble Space Telescope (HST) have
shown that many early-type galaxies have a large amount of dust
($10^{3}-10^{7} \Msun$), either in the form of a nuclear disk or in a
more diverse shape of filaments. Among different studies there is a
large variation in the detection rates which may be due to the
different methods, resolutions, and sensitivities of the observations
(Sadler \& Gerhard 1985 40\%; Goudfrooij et al. 1994 41\%; van
Dokkum \& Franx 1995 48\%; Ferrari \etal 1999 75\%; Tomita \etal
2000 56\%; Rest \etal 2001 43\%; Tran \etal 2001 (IRAS bright
sample) 78\%), but the general conclusion is that dust is common in
nearby ellipticals.

Establishing the presence of dust in nearby early type galaxies is
only the first step towards determining the role of dust in these
systems. It is already a well-known fact that radio-loud ellipticals
often have large amounts of dust but there are some open questions,
specially for the radio weak sources.  Verdoes Kleijn \etal (1999)
found that the incidence of dust in radio-loud early type galaxies is
89\% while Tran \etal (2001) has a value of 43\% for the occurrence of
dust in their snapshot sample of relatively radio-quiet nearby
early-type galaxies (for a description of the sample see Sec. 2). In
the same sample, 66\% of dusty galaxies have NRAO VLA Sky Survey
(NVSS) 1.4 GHz flux detections (Condon \etal 1998), while only 8\% of
galaxies without dust are listed as radio sources.

These results raise a question: how important is the presence of dust
for radio emission in the nuclei of ellipticals? Plausibly, dust
indicates the presence of gas, and gas is necessary to fuel the
activity of a central massive black hole (BH). However this line of
reasoning is highly incomplete. Gas may be present without dust.  Dust
may be present but not visually detectable (Goudfrooij \& de Jong
1995).  Dust and gas that have fed a BH in the past may not be
observable at the time when the nuclear activity is observed.  These
arguments justify a careful study of the relation between dust and
nuclear radio emission to determine the relevance of radio luminosity,
dust morphology and other effects.
 
There are two approaches to the study of extragalactic radio
sources. The first one is based on catalogs of discrete radio sources
followed by an analysis of the optical counterparts. The second
involves searching for radio emission from optically chosen
objects. The first approach (e.g.  de Koff \etal 2000), is
relatively efficient in finding radio galaxies, but emphasizes
powerful radio sources and may not provide a good counter sample of
radio quiet galaxies.  The second approach conversely emphasizes weak
radio sources (e.g. Sadler, Jenkins \& Kotanyi 1989; Wrobel 1991;
Wrobel \& Heeschen 1991; Sadler \etal 2002).

Both types of radio surveys are important.  Here we have chosen the
second method primarily so that the optical selection of the sample,
including Hubble type and especially dust content is not biased by
{\it a priori} selection for radio emission or other ``interesting"
properties of the galaxies.  The survey objects are selected on their
optical/IR properties only and then observed with the VLA with the
purpose of establishing the presence of nuclear AGNs. We compare our
dusty and non-dusty parts of the sample to investigate the importance of
dust (as a fuel reservoir) for the existence of nuclear activity.

In Section 2 we present the sample and discuss the observations and
the data reduction. In Section 3 we present the results of our
study. They are followed with a discussion in Section 4. Section
5 brings a discussion on correlation of dust with radio emission. The
conclusions are given in Section 6.

\section{Observations}

\subsection{Sample}

Our sample is compiled from two different samples described by Rest
\etal (2001) and Tran \etal (2001). The first sample
was created by selecting E/S0 galaxies on their optical properties
only from the Lyon/Meudon Extragalactic Database (LEDA). A randomly
selected subset of 68 galaxies from this sample were observed with HST
using WFPC2 in snapshot mode and thus this sample is referred to as
the ``snapshot'' sample. An additional sample of galaxies was
assembled from archival HST images of nearby E/S0 galaxies selected
for their 100 $\mu$m IRAS emission as these were likely to contain
large amount of dust (Tran \etal 2001). This sample is referred to as the
``IRAS sample''. From these two samples, we have selected 36 objects
(18 from each) according to their optical/IR properties, with no
regard to radio properties.  The reason for this selection was to
avoid biasing in picking a priori "interesting" objects and objects
with strong radio fluxes. Galaxies in our sample are nearby
ellipticals and lenticulars (E/S0), $cz<3200 \kms$, at galactic latitude
exceeding $20^{\circ}$ to minimize Galactic extinction, and with
absolute V-band magnitude less then -17.  Because of their optical/IR
selection they tend to have low radio powers.

The global properties of galaxies in our sample are listed in Table 1
of Rest \etal (2001) and in Table 7 of Tran \etal (2001). In the list
of 36 galaxies, 18 of them were chosen because they have dust in the
form of disks or filaments. The other 18 non-dusty galaxies were
selected to match dusty galaxies in optical properties, redshift,
magnitude, and IRAS flux. However, after the initial selection, more
detailed studies (Rest \etal 2001, Tran \etal 2001) showed that 6 of
the ``non-dusty'' galaxies showed faint dust structures and have here
been included in the ``dust'' class. We used H=80 $\kms\Mpc^{-1}$ to
be consistent with the papers defining the samples.

\subsection{Data Acquisition and Reduction}

The observations were undertaken with the VLA in C configuration at
3.6 cm wavelength. All sources were observed at two frequencies in the
8 GHz X-band (8.4351 and 8.4851 GHz) with a bandwidth of 50 MHz for
each frequency. We observed 68 sources in total, 36 galaxies and 32
calibrators. Each galaxy was observed for 15 minutes while calibrators
were observed for 130 seconds. Most of the calibrators had position
code A (positional accuracy $<$ 0\farcs002), but four calibrators had
B (0\farcs002 - 0\farcs01) and three had C (0\farcs01 - 0\farcs15) as
is indicated on the calibrator web page of the VLA. The radio
positions of the detected sources are limited by this positional
accuracy of the calibrators, as well as by the accuracy of the
Gaussian fit to the source brightness distribution, which is dependent
on the signal-to-noise ratios. Taking this in account the overall
accuracy is about 50 mas for mJy sources and about 100 mas for 100
$\mu$Jy sources. The observations were taken on March 13, 2000.

We used the Astronomical Image Processing System (AIPS) to reduce the
data using the standard procedures from the AIPS cookbook. After
initial calibration, the data were imaged using the task IMGR. The data
were self-calibrated in phases to improve the image dynamic range,
using a model derived from the same data. In some cases amplitude
self-calibration was performed on the data to improve the final
images. For our astrometric purpose, the positions of the sources were
extracted before self-calibration so that phase information was
preserved. All the images were examined using the tasks JMFIT and IMSTAT.

\section{Results}


\begin{table*}

 \caption[]{Radio properties of the galaxies. }
 \label{t:data}
\begin{tabular}{cccrllrrr}

\hline

\multicolumn{1}{c}{name}&
\multicolumn{1}{c}{dust}&
\multicolumn{1}{c}{D}&
\multicolumn{1}{c}{peak flux}&
\multicolumn{1}{c}{RA}&
\multicolumn{1}{c}{DEC}&
\multicolumn{1}{c}{L}&
\multicolumn{1}{c}{$\delta$}&
\multicolumn{1}{c}{NVSS flux}\\
\multicolumn{1}{c}{(1)}&
\multicolumn{1}{c}{(2)}&
\multicolumn{1}{c}{(3)}&
\multicolumn{1}{c}{(4)}&
\multicolumn{1}{c}{(5)}&
\multicolumn{1}{c}{(6)}&
\multicolumn{1}{c}{(7)}&
\multicolumn{1}{c}{(8)}&
\multicolumn{1}{c}{(9)}\\

\noalign{\smallskip}
\hline
\noalign{\smallskip}

ngc1400 & 2 & 25.4$^{P}$  & 2.092 $\pm$ 0.02 & 03 39 30.815  & -18 41 17.42 &  1.61 $\pm$ 0.02&1.80 & 2.5 $\pm$ 0.5 \\

ngc1439 &4 & 20.9$^{T}$ & $<$0.1 & & &&&\\

ngc2549 &0 & 15.7$^{R}$ &$<$0.1 & & & &&\\

ngc2592 &4 & 25.5$^{R}$ &0.41 $\pm$ 0.02 & 08 27 08.040 & 25 58 13.00 & 0.32 $\pm$ 0.01&0.65& \\

ngc2699 &4 & 21.8$^{R}$&$<$0.1 & & & &\\

ngc2768 &4 & 16.7$^{T}$ & 10.71 $\pm$ 0.02 &09 11 37.418 &60 02 14.84  &3.59 $\pm$ 0.01&0.54& 14.5 $\pm$ 0.6  \\

ngc2778 &0 & 25.4$^{T}$ & $<$0.1 & & &&\\

ngc2974 &3 & 25.9$^{T}$ & 5.22 $\pm$ 0.02 &09 42 33.310 &-03 41 57.09  & 4.19 $\pm$ 0.02&0.93 & 10.4 $\pm$ 0.5   \\

ngc2986* &0 & 22.3$^{T}$ & 8.40 $\pm$ 0.03 &09 44 27.256  &-21 16 11.23  &  &160.14&\\

ngc3078 &4 & 29.0$^{R}$ & 124.95 $\pm$ 0.04 &09 58 24.630  &-26 55 36.09 &125.73 $\pm$ 0.04&1.45 &279 $\pm$ 8\hspace{7.6pt}  \\

ESO437-15 &3 & 32.3$^{R}$ & 1.76 $\pm$ 0.04 &10 36 58.100 &-28 10 34.70 &2.20 $\pm$ 0.05&0.80 &3.2 $\pm$ 0.6\\

ngc3156 &2 & 14.0$^{T}$ & $<$0.1 & & & &&\\

ngc3226 &3 & 17.3$^{R}$ & 7.29 $\pm$ 0.05 &10 23 27.005  &19 53 54.75  &2.61 $\pm$ 0.02&0.97 &\\

ngc3348 &0 & 38.5$^{R}$ & 1.66 $\pm$ 0.02 & 10 47 10.000  &72 50 22.71 &2.94 $\pm$ 0.04&1.36& 7.8 $\pm$ 0.5\\

ngc3377 &1 & 9.1$^{T}$ & $<$0.1 & & && \\

ESO378-20 &0 & 35.6$^{R}$ & $<$0.1& & && \\

ngc3595 &0 & 30.4$^{R}$ & 0.22 $\pm$ 0.01 &11 15 25.180  &47 26 50.60  & 0.24 $\pm$ 0.01&3.87 &\\

ngc3610* &3 & 26.8$^{R}$ & 1.17 $\pm$ 0.03 &11 18 20.700  &58 49 38.11 &&230.78& \\

ngc4125* &4 & 20.1$^{R}$ & 1.23 $\pm$ 0.02 &12 08 04.180  &65 09 41.32&&86.29&\\

ngc4233 &4 & 29.6$^{R}$ & 2.52 $\pm$ 0.01 &12 17 07.679 &07 37 27.33  &2.64 $\pm$ 0.01&1.02&2.9 $\pm$ 0.5 \\

ngc4365 &0 & 15.7$^{R}$ & $<$0.1 & & & &&\\

ngc4406&4 & 17.0$^{V}$ & 0.59 $\pm$ 0.02 & 12 26 11.770 & 12 56 46.40 & 0.204 $\pm$ 0.07&1.37&  \\

ngc4476&3 & 24.7$^{T}$ & $<$0.1 & & &\\

ngc4494& 4& 17.8$^{R}$ & 0.27 $\pm$ 0.01&12 31 24.030  &25 46 30.01 &0.10 $\pm$ 0.01 &2.00& \\

ngc4552&4 & 17.0$^{V}$ & 93.40 $ \pm$ 0.02 &12 35 39.805  &12 33 22.78  &32.30 $\pm$ 0.01&0.35& 100 $\pm$ 3\hspace{7.6pt} \\

ngc4697&4 & 15.5$^{T}$ & $<$0.1 & & & &&\\

ngc4742&3& 15.9$^{T}$ & $<$0.1 & & &&&\\

ngc5198&0 & 34.1$^{R}$ & 0.83 $\pm$ 0.02 &13 30 11.390  &46 40 14.80 &1.15 $\pm$ 0.03 &1.16&3.6 $\pm$ 0.4 \\

ngc5322&4 & 23.9$^{T}$ & 13.60 $\pm$ 0.02 &13 49 15.269  &60 11 25.92  &9.33 $\pm$ 0.01&1.08 &64 $\pm$ 2\hspace{7.6pt} \\

ngc5557& 0& 42.5$^{R}$ & $<$0.1 & & & &&\\

ngc5576&0 & 19.1$^{R}$ & $<$0.1 & & & & &\\

ngc5812& 4& 24.6$^{R}$ & $<$0.1 & & &&&\\

ngc5813& 4& 24.6$^{R}$ & 2.95 $\pm$ 0.02 &15 01 11.234 & 01 42 07.10 &2.14 $\pm$0.01&0.72&12.3 $\pm$ 0.7\\

ngc5845& 4& 18.1$^{T}$ & $<$0.1 & & & &&\\

ngc5982&0 & 39.3$^{R}$ & $<$0.1 & & &&&\\

ngc6278& 0& 37.1$^{R}$ & 1.06 $\pm$ 0.01 &17 00 50.325  &23 00 39.73  &1.75 $\pm$ 0.02&0.62& \\

\noalign{\smallskip}

\hline

\end{tabular}\\

{ Col.(1): name of the galaxy, sources with a star are far from the
 nuclear region of the corresponding galaxies and were treated as non
 detections; Col.(2): level of dust: 0 = no dust, 1 = filamentary low,
 2 = filamentary medium, 3 = filamentary high, 4 = dusty disk (Tran
 \etal 2001); Col.(3): distance in Mpc from (P) - Perrett \etal 1997,
 (T) - Tran \etal 2001, (R) - Rest \etal 2001, (V) - Virgo galaxies,
 assumed to be at distance of 17 Mpc; Col.(4): flux at 3.6cm in mJy,
 or 4$\sigma$ upper limits for nondetections; Cols.(5) and (6): radio
 position (h, m, s) and (deg, arcmin, arcsec) from our maps
 (J2000); Col.(7): luminosity in $10^{20}$ WHz$^{-1}$; Col.(8): offset
 in arcseconds, between the 3.6 cm radio position and the position of
 the galaxy optical nucleus on HST images (Tran \etal 2001, Rest \etal
 2001); Col.(9): peak flux from NVSS survey (Condon \etal 1998)}

\end{table*}

Twenty galaxies in our sample of 36 were detected as radio
sources. Three detected sources (associated with NGC 2986, NGC 3610,
NGC 4125) can not be matched with the central regions of the galaxies
and there are no visible counterparts on the available HST
pictures, hence they are most likely background sources. The radio
sources lay far from the nuclei (about $2\farcm 67\!$ for NGC 2986,
$3\farcm84\!$ for NGC 3160, and $1\farcm44 \!$ for NGC 4125).
Although the fluxes and positions of these sources are listed in Table
1 (with asterisks) we treat them as {\it non-detections} of central
AGNs in the surveyed galaxies.  This leaves 17 AGN detections in 36
galaxies (47 \% detection rate). The smallest signal to noise ratio
(SNR) is about 10$\sigma$ with a survey average rms $\sigma=2.8 \times
10^{-5}\Jybeam$. For non-detected sources we calculated 4$\sigma$
upper limits on detection, thus, the detection limit of our survey is
about $0.1$ mJy. Radio properties of the sample are given in Table 1.
By comparison the detection limit of the NVSS (Condon \etal 1998) used
by Tran \etal (2001) to discuss radio properties of our sample is
$\sim 3$ mJy, a factor of 30 higher.

Most of the detections are point-like, unresolved structures. NGC 5322
is the only galaxy with noticeable jet-like structure. Typical
detected sources are on the level of a few mJy; the weakest detections
were $\sim 200 \mu$Jy.  Of the 36 galaxies in the sample, 24 galaxies show
disk or filamentary dust structure and 13 (54\%) of them are detected
as radio sources. Twelve show no dust of which four (33\%) are
detected.

\section{Discussion}

\subsection{Nature of Detected Radio Sources}

Most of the detections are unresolved radio sources easily associated
with the central $1\farcs5$ on the HST image. At $25 \Mpc$, the mean
distance of the galaxies in the sample, $1\arcsec$ is about 120
pc. Thus the emission is clearly (near) nuclear, but not necessarily
of AGN origin.  Since the sources are weak (radio power ranges from
$10^{19}$ W Hz$^{-1}$ to $10^{21}$ W Hz$^{-1}$ with a few higher
exceptions) there is a possibility that they arise from a non-AGN
mechanism, e.g. nuclear starbursts. Since we are interested in the
AGN/dust connection we wish to exclude this possibility.  We argue
here that the dominant source of radio emission in our detections is a
non-thermal mechanism similar to that which operates in more powerful
radio sources.

There are several radio and infrared criteria that can be used to
distinguish between emission from starburst and AGN galaxies: (i)
radio mor\-pho\-lo\-gy, (ii) far-infrared to radio flux-density parameter \(
u\equiv \log(S_{\rm 60\mu m}/S_{\rm 1.4 GHz}) \), (iii) infrared spectral
index $ \alpha_{IR}$ \(\equiv \log(S_{\rm 60\mu m}/S_{\rm 25\mu
m})/\log(60/25) \) (Condon \& Broderick 1988 and 1991; Condon \etal
1991; Condon, Huang \& Thuan, 1991), and (iv) the steepness of the
radio spectra.  Radio morphology implies coherent radio jets and radio
lobes that may lie well outside the optical galaxy. Starburst galaxies
usually have $u\geq1.6$, and $\alpha_{IR}\geq +1.25$. Steepness of the
radio spectrum is also used as a criterion since optically thick
AGN cores usually have flat spectra, while the dominant emission 
from star-forming regions (supernova remnants, and cosmic rays diffusing 
from them) have steep spectra. Nearly all
spirals and unclassifiable objects (e.g mergers) have steep
spectra, while flat spectra and other AGN characteristics (radio
morphology, $u\leq1.6$, and $\alpha_{IR}\leq +1.25$) are associated
with ellipticals (Sadler, Jenkins, \& Kotanyi 1989, Condon 1991).

All detected galaxies in our sample have low-luminosity unresolved
sources in the innermost central regions. Although the sources are
certainly nuclear in origin (suggesting AGN activity) any radio
classification according to radio morphology is not possible (except
in the clear case of a jet in NGC 5233). Half of the galaxies were
picked based on their large scale dust and infrared properties from
IRAS survey. This means that those galaxies are going to have larger
$\alpha_{IR}$ indices, which would mark them as starburst, although
they still might have nuclear AGN which are the subject under
discussion. Using the large scale IR emission to determine the nature
of the nuclear radio emission doesn't seem to be a very good
discriminator between SBs and AGNs. However, most of our objects have
measured nuclear H$\alpha$ fluxes or upper limits (Tran, private
communication), and standard calculations (Osterbrock 1989) indicate
that the free-free fluxes from these regions would be below 3 $\mu$Jy,
which is about two to three orders of magnitude smaller than
our observed fluxes. Other evidence that we are dealing with
non-thermal radiation comes from the flatness of the spectra in our
sample. Eleven of the galaxies were detected before in the NVSS
(Condon \etal 1998) and comparing the fluxes at our frequency (8.45
GHz) and the frequency of the NVSS (1.4 GHz) it is clear that most of
the detected galaxies have flat spectra (Table 1).

Previous studies (Phillips \etal 1986; Sadler, Jenkins \& Kotanyi
1989) have shown that HII regions in early type galaxies are
not likely to contribute to the radio galaxy population above
$10^{19}$ W Hz$^{-1}$.  Keeping in mind that all galaxies in our
sample are Es and S0s, that emission is confined to nuclei of the host
galaxies, and that the sources have flat spectra, we can assume that
the dominant radio component in our case is synchrotron emission from
an active nucleus producing low-luminosity counterpart of more
distant, luminous AGNs.

\subsection{Radio Luminosity Function}


 \begin{figure}
   \centering
   \includegraphics[width=\columnwidth]{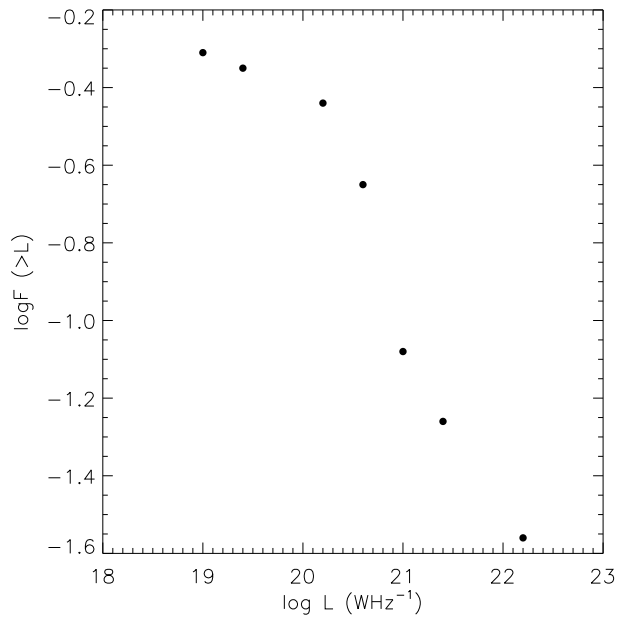}   
   \caption{ 
The integral luminosity function derived from 17 detected sources out
of a total of 36.  The dots represent a crude LF calculated from the
detections as the integral of a series of delta functions. A detection
at L$_i$ contributes 1/N${_d(L_{i})}$ where the denominator is the
number surveyed galaxies detectable at L$_i$.  The steep rise flattens
off below of $10^{20}$W Hz$^{-1}$. The error bars are not plotted
since the bins in the integral RLF are not independent.}
   \label{Fig1}%
    \end{figure}


\begin{figure}
   \centering
   \includegraphics[width=\columnwidth]{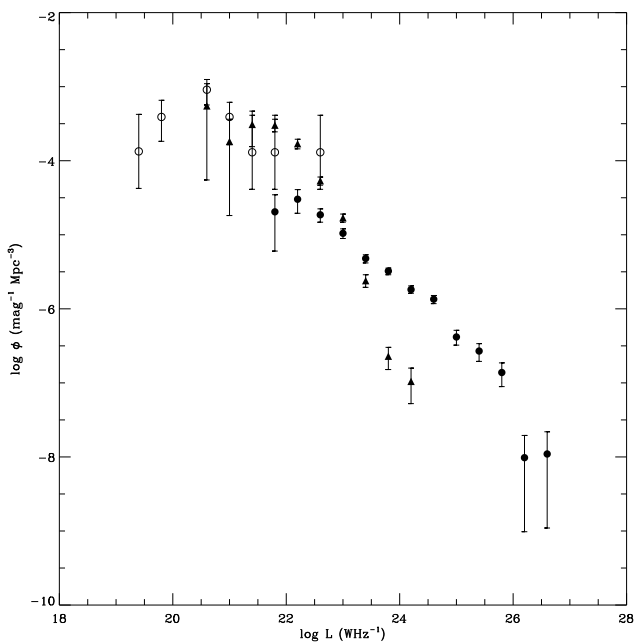}
   \caption{
Comparison of AGN and starburst (SB) local radio luminosity
function. Filled symbols (circles: AGN, triangles: SB) are data from
Sadler \etal (2002) and Condon (1991), while open circles are our
data. The local density of AGN rises continuously at low luminosities,
reaching the value of SB, suggesting that AGNs are as common as SB in
local universe. It is possible that at this low luminosity level both
processes are present in galaxies, but in some galaxies one of the
engines is stronger.}
   \label{Fig3}%
\end{figure}

The size of our sample is too small and too limited in radio
luminosity range to construct a complete local radio luminosity
function (RLF) of early-type galaxies. In any case, the sample was not
constructed for that purpose. Still, we can make a useful estimate of
the low luminosity end of the local RLF in order to see how
it corresponds with previously found local RLFs and offer an estimate
of the behavior of RLF at low luminosities. For this purpose we define
the fractional luminosity function (Auriemma \etal 1977):
\begin{equation}
  F_i(L,z) = \rho_i(L, z)/\varphi_i(z),
\label{e:lumdef}
\end{equation}
where $\varphi_{i}(z)$ is the volume density of objects of type $i$ at
the redshift $z$, and $\rho_{i}(L, z)$ is the density of sources
associated with optical objects of type $i$ with the given radio
luminosity $L$ and at the given redshift $z$. The fraction of all
elliptical galaxies with luminosity at given frequency between $L$ and
$L+dL$ at the redshift $z$, of optical magnitude $M$ is given then by
$F_{E,M}(L,z)dL$. In order to estimate the ``bivariate'' RLF defined
in this way we can calculate the fractional detection
$f_{ij}=n_{ij}/N_{ij}$, where $n(L_i, M_j)$ is the number of actually
detected galaxies within the optical magnitude range $M_j\pm0.5$ and
radio luminosity interval log$L_i\pm0.2$, while $N(L_i, M_j)$ is the
number of galaxies in the sample which could have been detected if
their optical magnitude and radio luminosity were in the given
interval. In our case we did not bin in optical magnitude but only in
radio power since the sample is limited. Our first estimate of the RLF
is then given by $f_L=n_L/N_L$ and it is shown in Figure 1 in integral
form, F($>$L).  As it can be seen from the Figure 1, the integral RLF
of nearby ellipticals rises steeply with decreasing of the radio
luminosity and only at the lowest intensities ($10^{19}$ W Hz$^{-1}$)
levels off at a point where $\sim 50$\% of all E/S0 galaxies show
activity.

Previous RLFs (Auriemma \etal 1977; Sadler, Jenkins, \& Kotanyi 1989;
Condon 1991; Sadler \etal 2002) of nearby ellipticals with AGN
signature were made for galaxies with radio luminosities higher than
$10^{21} - 10^{22}$ W Hz$^{-1}$. The more recent studies considered
also starburst galaxies. While AGN were found in ellipticals,
starbursts inhabited spirals. These different distributions had
different RLFs and often starburst RLFs extended to the level of
$10^{20}$W Hz$^{-1}$. With our low luminosity data, we are able to
extend the existing RLFs of AGN down to $10^{19}$W Hz$^{-1}$ and can
construct an exclusively AGN RLF.

We compare our data with two studies (Sadler, Jenkins, \& Kotanyi 1989
and Condon 1991) in Figure 2 (AGNs and starbursts plotted). We
have converted our differential data from $F(L)$ to a spatial density
$\phi$ (number of sources per Mpc$^3$ per 0.4 in log L) using the
value for spatial density of early type galaxies, from Sadler,
Jenkins, \& Kotanyi (1989), which is 10$^{-2.33}$ mag$^{-1}$
Mpc$^{-3}$. Gratifyingly our data agrees quite well with the previous
data in the region of overlap.  Together these data confirm the
flattening of the RLF for AGNs below $\sim 10^{20}$ W Hz$^{-1}$.  It
is also interesting that the space density of low luminosity AGN is
very similar to starbursts galaxies of the same luminosities. The RLFs
of the two distributions are basically overlapping in this luminosity
range.

\section{Correlation of dust with radio emission}

\subsection{Crude Statistics}


\begin{figure}
   \centering
   \includegraphics[width=\columnwidth]{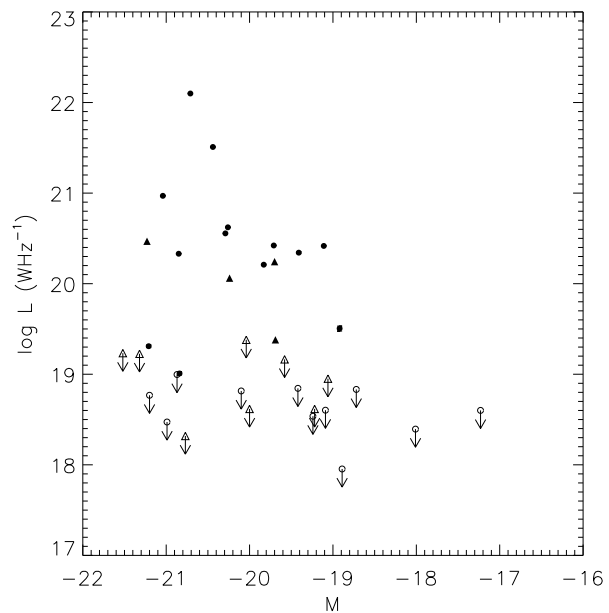}
   \caption{
Plot of log radio luminosity in W Hz$^{-1}$ versus absolute optical
magnitude. Filled symbols are radio-detected nuclei of galaxies,
while open symbols indicate upper limits for the rest of the
galaxies. Triangles are sources in galaxies without dust and circles
are sources in dusty galaxies. Sources in dusty galaxies have a slight
tendency for being more powerful than the sources in non-dusty
galaxies. Luminosity error bars are smaller than the symbols}
   \label{Fig4}
\end{figure}

The HST pictures of the galaxies in the original sample (Tran \etal
2001) confirm that dust is very common in ellipticals.  There are two
different morphologies in which dust appears in the galaxies from our
sample: disky and filamentary. We have 15 galaxies with disks
and 9 with large amount of dust in filaments. Thirteen of the 24 dusty
galaxies have a radio detection (54\%), while 4 out of 12 non-dusty
galaxies show a detection (33\%). There is no significant difference
in radio luminosity between the galaxies with disky and filamentary
dust: 60\% detections in galaxies with disks and
44\% in galaxies with filaments. This finding is in general
agreement with the findings by Tran \etal (2001).

The relationship between optical absolute magnitude and radio
luminosity for our weak radio sources is shown in Figure 3. There is
little difference in the distributions of the dusty and non-dusty
galaxies, except perhaps that the three most powerful galaxies are all
dusty. As expected, the more powerful radio sources are found in the
brighter galaxies.

Most nearby high luminosity radio sources are found in dusty early
type galaxies (de Koff et al.(2000)), which suggests a link between
dust and the existence of a radio source. Our wish now is to see if at
the lower levels of radio luminosity dust also plays an important
role. We divide the detections in two sets of sources: dusty and
non-dusty according to the descriptions in Tran \etal (2001). As we
see above, the dusty galaxies show a somewhat higher detection rate,
but, given the steepness of the RLF, this could be influenced by
slight differences in the distances to the two samples, or slight
differences in the achieved sensitivities.  Therefore it is more
meaningful to compare the RLFs of the two samples than the detection
percentages.

\subsection{Comparison of RLFs}


\begin{figure}
   \centering
   \includegraphics[width=\columnwidth]{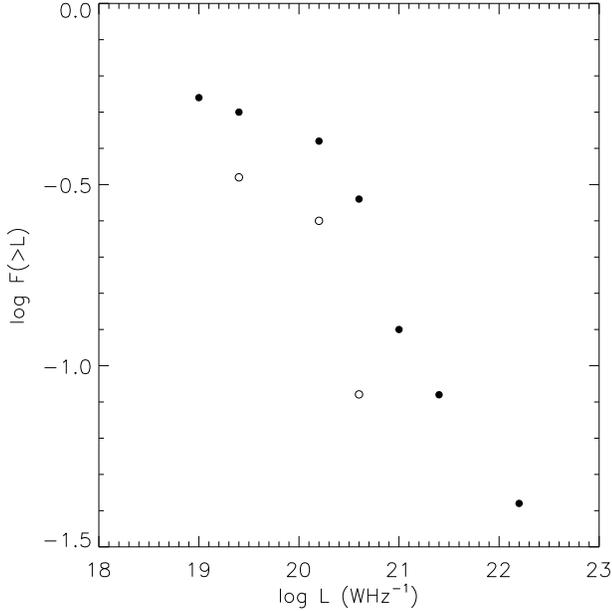}
   \caption{
The separated integral luminosity function. Open circles present the
RLF for sources from galaxies lacking dust. Filled circles present the
RLF of sources from dusty galaxies. Statistical tests show that the
two distributions are not distinguishable, suggesting that dust is not
important for the existence of low-luminosity AGN in nearby early-type
galaxies}
\label{Fig5}%
    \end{figure}


\begin{figure}
   \centering
   \includegraphics[width=\columnwidth]{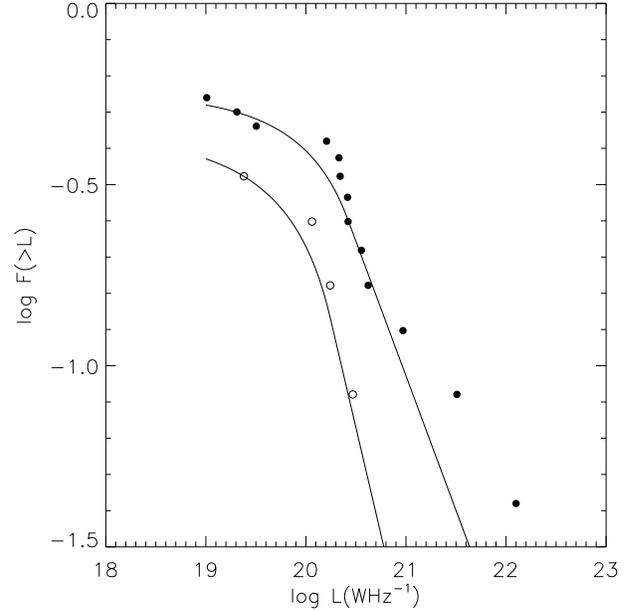}
   \caption{ Estimated integral luminosity functions compared to the
data.  The drawn lines show the ML model fits (equations 2 and 3) to
the data, while the points are computed from individual
detections. Field circles present dusty galaxies while open circles
present non-dusty galaxies. The thick line shows the ML model fit for
dusty galaxies and the thin line for non-dusty galaxies.}
   \label{Fig6}%
    \end{figure}


\begin{table*}

 \caption[]{Estimated integral luminosty coefficients. Values of coefficeints of estemated integral luminosity function F obtained by maximum likelihood method. Errors are 1$\sigma$ estimates.}
  \label{t:mlm}
$$
  \begin{array}{ccccc}
    \hline
    \noalign{\smallskip}
    $data sample$ & \alpha & \beta & x_{c} &  A_{0}\\
    \noalign{\smallskip}
    \hline
    \noalign{\smallskip}
    $all data$\hspace{10pt} & -0.6 \pm 0.3\hspace{10pt} & 0.80 \pm 0.10\hspace{10pt}& 20.41 \pm 0.04\hspace{10pt} & 0.21 \pm 0.04\\ 
    $dust data$\hspace{10pt} & -0.6 \pm 0.4\hspace{10pt} & 0.74 \pm 0.09\hspace{10pt}& 20.40 \pm 0.10\hspace{10pt} & 0.25 \pm 0.07 \\ 
    $non-dust data$\hspace{10pt} & -0.5 \pm 0.3\hspace{10pt} & 1.20 \pm 0.60\hspace{10pt} & 20.24 \pm 0.04\hspace{10pt} & 0.14 \pm 0.02\\
    \noalign{\smallskip}
    \hline
  \end{array}
$$

\end{table*}

The integral RLFs for the two samples, computed by the
same algorithm as that in Fig. 1 for the whole sample,
are shown in Figure 4. In this representation also, the
dusty galaxies seem more active, but the difference
is relatively small (a factor of $\sim 1.6$) and we
wish to test the significance of this difference.

We have tried two statistical tests: Kolmogorov-Smirnoff (K-S) and a
test using maximum likelihood method (ML). The K-S test has the advantage of
being parameter and form free, but the disadvantage of not being very
conclusive for small samples. We have two data sets,
one with 13 sources in dusty galaxies and one with 4 sources in
non-dusty galaxies. We used routines from Numerical Recipes (Press
\etal 1992). The probability that these two observation sets could be
obtained from the ${\it same}$ RLF is 64\%, hence the RLFs are
statistically indistinguishable. However, the K-S test is sensitive to the
effective number of data points, N$_e$, which in our case of two
distributions is N$_e$ = N$_1$N$_2$/(N$_1$ + N$_2$) = 3.1. Press \etal
(1992) give N$_e \ge 4$ as a limit for a decent accuracy. Thus, the
above probability is not very accurate, but it still
implies that the two data sets (two luminosity functions) are not
significantly different.

Another approach that is more sensitive, but requires more {\it a
priori} assumptions, is to fit a specific, parameterized, function to
the RLF data using the maximum likelihood method, and compare the
fitted functions. Since there are a limited number of degrees of
freedom in the function, more powerful statistical statements can be
made. To estimate the integral luminosity function we used a set of
two power-law functions allowing for a break in the RLF. Our choice is
similar to some previously used functions (Auriemma \etal 1977):

\begin{equation}
F = A_{0} \cdot 10^{-\beta(x-x_{c})} {\rm\hspace{3.3cm} for}\; x>x_{c}
\end{equation}
\begin{equation}
F = A_{0} \cdot ( 1+\frac{\beta}{\alpha}( 10^{-\alpha(x-x_{c})} - 1))
{\rm\hspace{1cm} for}\; x<x_{c}
\label{e:lum}
\end{equation}

\noindent where x = log$_{10}$L, and L is radio luminosity in W Hz$^{-1}$. The
normalization constant A$_{0}$ is chosen so that at x = x$_{c}$, F =
A$_{0}$.  Originally we assumed that at low luminosities the value of $F(x)$
had to approach $F=1$ as $x\rightarrow -\infty$, thus providing an additional
constraint on the model.  These solutions provided poor fits to the data and
were dropped.  This implies, however, that there is another break (or
continuous change of slope) in the RLFs below the limits of our survey.

The system of coefficients $\alpha$, $\beta$ (slopes of the curve),
x$_{c}$ (the position of the break) and A$_{0}$ (normalization) that
maximize the probability in the method, provide also the best fit to
the data. Table 2 contains the calculated values for $\alpha$,
$\beta$, x$_{c}$ and A$_{0}$.

The best-fit integral luminosity functions are compared to the
observed values in Figure 5. The symbols are filled circles for dusty and
open circles for non-dusty sources. The thick represents the model fit
to dusty sources, while the thin line shows the fit to the non-dusty.
The model curves fit the individual points reasonably well.  The three
radio brightest galaxies lie somewhat above the best two power-law fit
in the region of the break, but the best fit value of the slope above
the break, $\beta\simeq 0.75$, agrees with the slope measured by
Sadler \etal (2002) (Fig. 2) based on much more data in the higher
luminosity ranges.

The ML parameters $\alpha$, $\beta$, and $x_c$ are essentially
identical for dusty and non-dusty galaxies, indicating that
the forms of the RLF are similar.  Not surprisingly the normalization
$A_0$ is higher for the dusty galaxies by a factor of about 1.8,
but this is only 1.6 times the uncertainty.

Another way to globally judge the significance of the difference
between these RLFs it to ask if the ``true" RLF were given by the
dusty model, how unlikely is it that we would only detect four (or
less) of the twelve non-dusty galaxies.  If this probability is small,
then the samples are significantly different. From Poisson statistics
the probability of 4 or less non-dusty detections given the dusty RLF
(thick line on Fig. 5 or the second line values in the Table 2) is
27\%, indicating a low statistical significance.  If, hypothetically,
the true non-dusty RLF is a factor of 1.6 lower than the dusty RLF we
can ask how many non-dusty galaxies must be surveyed in order to
demonstrate the RLF difference at a reliability of, say, 5\%.
Repeating the Poisson analysis indicates that a sample about four
times the current size is needed, or about 50 non-dusty galaxies.

Perhaps the most important result of this investigation is that
in any case, a sizable fraction of the non-dusty galaxies, $\sim 30$\%,
are radio emitters, so that the presence of visible dust
is {\it not necessary} for radio emission from an AGN.

\section{Conclusions}

We report 3.6 cm VLA observations of a sample of 36 near-by
ellipticals selected on their optical/IR properties. We detected 17
unresolved (except the jet in NGC5322), compact, flat-spectrum radio cores
associated with the central 1\arcsec $\;$ of the nuclei, suggesting
that all detected sources are low luminosity AGNs. The lowest detected
luminosities are $\sim 10^{19}$ W Hz$^{-1}$.

We determine the Radio Luminosity Function (RLF) from these galaxies
down to a luminosity almost two orders of magnitude lower in
luminosity than previously published studies. It shows the
continuation in the rise of space density of sources with AGN
signature, which was expected from other, unpublished, studies
(Condon, private communication). At the luminosities considered
(i.e L $\sim$ 10$^{19}$ - 10$^{22}$ WHz$^{-1}$), the space
densities of the AGNs and starburst galaxies approach each other,
becoming hardly distinguishable. At the lower luminosity end of our
sample $\sim 50$\% of E/S0 galaxies have detectable radio-AGNs.

Although the non-dusty galaxies show an indication of a lower
probability of radio emission, the difference is not statistically
significant in a sample of this size.  Dust detectable in HST images
is certainly not necessary for nuclear radio emission. This situation
may be different for the more powerful radio galaxies observed in
earlier surveys.

This takes us back to the question of fuel for the central engine of
our low luminosity AGNs. If fuel is necessary for nuclear activity why
do we find weak AGNs without visible dust? It should be noted that our
non-dusty galaxies with radio detections lay further away (D $>$ 30
Mpc) and it might be possible that small amounts of dust were not
detected. Also, extremely diffuse gas and dust would not be visible
(Goudfrooij \& de Jong, 1995) but current theories of accretion
require bars and disks and other distinct structures, so fueling from
diffuse gas seems unlikely. Similarly, these galaxies could be fueled
by hot, dust-free gas, but it seems unlikely that any mechanism in
these low-luminosity sources would destroy dust more than in the high
luminosity sources were dust is common. A more likely explanation is
that the amount of dust and gas present near the nucleus is in some
sense positively correlated to the AGN luminosity.  The sources in our
study are two to three orders of magnitude less luminous than the 3C
sources in de Koff \etal (2000), where typical dust optical depths
were of order unity. In the HST images, optical depths of less than
1\% would probably be missed.  Alternatively, AGN fueling may be
cyclic, and AGN radio emission is now fueled from material at a few
Schwarzschild radii, after the material in a larger circumnuclear
accretion disk has been temporarily consumed.

The luminosity of an AGN is determined by the fueling rate and the
mass-to-radiation conversion efficiency. The latter is influenced by
the degree of advection which is in turn influenced by the Eddington
luminosity and the mass of the BH. As recent evidence suggests (Ho,
2002 and references therein) many of the characteristics of low
luminosity AGN could be explained by an advection dominated
accretion flow (Narayan \& Yi 1995; Narayan, Mahadevan, \&
Quataert 1998). The explanation of the dust/radio emission/luminosity
relations may perhaps be found when we know the BH masses of the
galaxies, or when we understand the characteristics of non-steady
accretion flows.

\begin{acknowledgements}
The VLA is operated by the National Radio Astronomy Observatory for the U.S.
National Science Foundation.
\end{acknowledgements}



\begin{thebibliography}{}

\bibitem[Auriemma et al.(1977)]{1977A&A....57...41A} Auriemma, C., Perola, 
G.~C., Ekers, R.~D., Fanti, R., Lari, C., Jaffe, W.~J., \& Ulrich,
M.~H.\ 1977, \aap, 57, 41

\bibitem[]{}
Goudfrooij, P., Hansen L., J{\o}rgensen, H. E., \& N{\o}rgaard-Nielsen, H. U. 
1994, A\&AS, 105, 341

\bibitem[Goudfrooij \& de Jong (1995)]{1995A&A....298...784G}
Groudfrooij, P., de Jong, T.\ 1995, \aap, 298, 784

\bibitem[Condon et al.(1998)]{1998AJ....115.1693C} Condon, J.~J., Cotton, 
W.~D., Greisen, E.~W., Yin, Q.~F., Perley, R.~A., Taylor, G.~B., \& 
Broderick, J.~J.\ 1998, \aj, 115, 1693 

\bibitem[Condon \& Broderick(1988)]{1988AJ.....96.30C} Condon, J.~J.~\& 
Broderick, J.~J.\ 1988, \aj, 96, 30 

\bibitem[Condon \& Broderick(1991)]{1991AJ....102...1663C} Condon, J.~J.~\& 
Broderick, J.~J. 1991, \aj, 102, 1663

\bibitem[Condon(1991)]{1991osgl.work..113C} Condon, J.~J.\ 1991, ASP 
Conf.~Ser.~18: The Interpretation of Modern Synthesis Observations of 
Spiral Galaxies, 113 

\bibitem[Condon, Huang, Yin, \& Thuan(1991)]{1991ApJ...378...65C} Condon, 
J.~J., Huang, Z.-P., Yin, Q.~F., \& Thuan, T.~X.\ 1991, \apj, 378, 65 

\bibitem[Ferrari et al. (1999)]{1999A&AS..136..269F}
Ferrari, F., Pastoriza, M.G., Macchetto, F., Caon, N. 1999, A\&AS, 136, 269

\bibitem[Ho, (2002)]{2002ApJ...564..120H} Ho, L.~C., 2002, \apj, 564, 120

\bibitem[de Koff et al.(2000)]{2000ApJS..129...33D} de Koff, S., Best, P., Baum, S.~A., Sparks, W., 
        R{\" o}ttgering, H., Miley, G., Golombek, D., 
        Macchetto, F., \&Martel, A., 2000, \apjs, 129, 33

\bibitem[Narayan, Mahadevan, \& Quataert(1998)]{1998tbha.conf..148N} 
Narayan, R., Mahadevan, R., \& Quataert, E.\ 1998, Theory of Black Hole 
Accretion Disks, 148 

\bibitem[Narayan \& Yi(1995)]{1995ApJ...452..710N} Narayan, R.~\& Yi, I.\ 
1995, \apj, 452, 710 

\bibitem[Osterbrock (1989)]{1989agna.book.....O} Osterbrock, D.E., 1989 {\it Astrophysics of gaseous
nebulae and active galactic nuclei}, University Science Books

\bibitem[Perrett et al.(1997)]{1997AJ....113..895P} Perrett, K.~M., Hanes, 
D.~A., Butterworth, S.~T., Kavelaars, J., Geisler, D. \& Harris, W.~E. 
1997 \textbf{\aj}, 113, 895


\bibitem[Phillips et al. (1986)]{1986AJ.....92..503P} Phillips, M.~M., Jenkins, C.~R., Dopita, M.~A., 
        Sadler, E.~M., \& Binette, L., 1986 \aj, 92, 503


\bibitem[Press et al. (1992)]{1992nrfa.book.....P}  Press, W.~H., Teukolsky, S.~A., Vetterling, W.~T., Flannery, B.~P., 1992, Numerical recipes: The Art of Scientific Computing, CUP, Cambridge

\bibitem[Rest et al.(2001)]{2001AJ....121.2431R} Rest, A., van den Bosch, 
F.~C., Jaffe, W., Tran, H., Tsvetanov, Z., Ford, H.~C., Davies, J., \& 
Schafer, J.\ 2001, \aj, 121, 2431 

\bibitem[Sadler \& Gerhard(1985)]{1985MNRAS.214..177S} Sadler, E.~M.~\& 
Gerhard, O.~E.\ 1985, \mnras, 214, 177

\bibitem[Sadler, Jenkins, \& Kotanyi(1989)]{1989MNRAS.240..591S} Sadler, 
E.~M., Jenkins, C.~R., \& Kotanyi, C.~G.\ 1989, \mnras, 240, 591 

\bibitem[Sadler et al. (2002)]{2002MNRAS.329..227S} Sadler, E.~M., Jackson, C.~A., Cannon, R.~D., 
        McIntyre, V.~J., Murphy, T., Bland-Hawthorn, J., 
        Bridges, T., Cole, S., Colless, M., Collins, C., 
        Couch, W. and, Dalton, G. and, De Propris, R., Driver, S.~P., 
        Efstathiou, G., Ellis, R.~S., Frenk, C.~S., Glazebrook, K. and 
        Lahav, O., Lewis, I., Lumsden, S., Maddox, S. and 
        Madgwick, D., Norberg, P., Peacock, J.~A.,Peterson, B.~A. and 
        Sutherland, W., \& Taylor, K., 2002, \mnras, 329, 227

\bibitem[]{}
Tomita, A., Aoki, K., Watanabe, M., Takata, T., \& Ichikawa, S. 2000, AJ, 120, 123

\bibitem[Tran et al.(2001)]{2001AJ....121.2928T} Tran, H.~D., Tsvetanov, 
Z., Ford, H.~C., Davies, J., Jaffe, W., van den Bosch, F.~C., \& Rest, A.\ 
2001, \aj, 121, 2928 

\bibitem[]{}
van Dokkum, P.G., \& Franx, M. 1995, AJ, 110, 2027
 
\bibitem[]{}
Verdoes Kleijn, G. A., Baum, S. A., de Zeeuw, P.T., \& O'Dea, C. P. 1999, AJ, 118, 2592 


\bibitem[Wrobel(1991)]{1991AJ....101..127W} Wrobel, J.~M.\ 1991, \aj, 101, 
127 

\bibitem[Wrobel \& Heeschen(1991)]{1991AJ....101..148W} Wrobel, J.~M.~\& 
Heeschen, D.~S.\ 1991, \aj, 101, 148 

\end{thebibliography}
\end{document}